\documentclass[12pt]{article}
\usepackage{amsmath}
\usepackage{amssymb}
\usepackage{latexsym}
\newtheorem{theorem}{Theorem}
\newtheorem{lemma}{Lemma}

\newtheorem{convention}{Convention}
\newtheorem{remark}{Remark}

\numberwithin{equation}{section}

\begin{document}
\date{}
\author{M.I.~Belishev
                   and A.F.~Vakulenko}
\title{$s$-points in $3\rm d$ acoustical scattering} \maketitle
\begin{abstract}
The notion of $s$-points has been introduced by the authors (SIAM
JMA, 39 (2008), 1821--1850) in connection with the control problem
for the dynamical system governed by the $3\rm d$ acoustical
equation $u_{tt}-\Delta u+qu=0$ with a real potential $q \in
C^\infty_0({{\mathbb R}^3})$ and controlled by incoming spherical waves. In
the generic case, this system is controllable in the relevant
sense, whereas $a \in {\mathbb R}^3$ is called a {\it $s$-point}
(we write $a \in \Upsilon_q$) if the system with the shifted
potential $q_a=q(\,\cdot-a)$ {\it is not controllable}. Such a
lack of controllability is related to the subtle physical effect:
in the system with the potential $q_a$ there exist the finite
energy waves vanishing in the past and future cones
simultaneously. The subject of the paper is the set $\Upsilon_q$:
we reveal its relation to the factorization of the $S$-matrix,
connections with the discrete spectrum of the Schr$\ddot{\rm
o}$dinger operator $-\Delta+q$ and the jet degeneration of the
polynomially growing solutions to the equation ${\left(-\Delta+q\right)} p=0$.
\end{abstract}
\setcounter{section}{-1}

\section{Introduction}
\subsection{Dynamical system}
An acoustical scattering problem is the system of the form
\begin{align}
\label{1} &u_{tt}-\Delta u+qu=0, \qquad (x,t) \in {{{\mathbb R}^3}} \times (-\infty,\infty) \\
\label{2} &u \mid_{|x|<-t} =0 , \qquad t<0\\
\label{3} &\lim_{s \to \infty}
su((s+\tau)\omega,-s)=f(\tau,\omega), \qquad (\tau,\omega)   \in
[0,\infty)\times S^2\,,
\end{align}
where $u=u^f(x,t)$ is a solution ({\it wave}), $q=q(x)$ is a real
valued smooth\footnote{everywhere in the parer "smooth" means
$C^{\infty}$-smooth.} compactly supported function ({\it
potential}), and $f \in {\cal F}
:=L_2\left([0,\infty);L_2\left(S^2\right)\right)$ is a {\it
control}. With the system one associates the {\it control
operator} $W: {{\cal F}} \to {{\cal H}} :=L_2({{\mathbb R}^3}), \,\,Wf:=u^f(\,\cdot\,,0)$ and
the subspaces: ${{\cal U}}:={\rm Ran\,}W$ ({\it reachable set}), ${{\cal D}}:={{\cal H}}
\ominus {{\cal U}}$ ({\it defect subspace}), ${{\cal N}}:={\rm Ker\,}W \subset {{\cal F}}$
({\it null control subspace}). The relations $0 \leq {\rm dim\,}{{\cal D}}
={\rm dim\,}{{\cal N}} < \infty$ hold \cite{BV3}. If ${{\cal D}}=\{0\}$, the
system (\ref{1})--(\ref{3}) is said to be controllable
\footnote{Note that the unperturbed system (with $q=0$) is
controllable.}; the case of ${{\cal D}} \not=\{0\}$ is referred to as a
lack of controllability.

\subsection{$s$-points}
As is shown in \cite{BV3}, the case ${\rm dim\,}{{\cal D}} \not=0$ is
realizable and corresponds to the curious effect: for $f \in {{\cal N}}$,
the function $w^f(x,t):=\int^t_0u^f(x,s)\,ds$ is a finite energy
solution of (\ref{1}), which satisfies $w^f(\,\cdot\,,
-t)=w^f(\,\cdot\,,t)$ and \begin{equation}\label{cones} w^f
\mid_{\,|x|<|t|}\,=\,0\,,\end{equation} i.e., describes the wave
vanishing in the past and future cones simultaneously. This wave
comes from infinity, stops at the moment $t=0$ \footnote{We mean
$w^f_t(\,\cdot\,,0)=0$.}, and then return back to infinity along
the same trajectory. Such a behavior motivates to call it a {\it
reversing wave} and regard the coordinate system origin $x=0$ as a
point, which is able to stop incoming waves ('stop point').
Looking for such points in the space, we change the coordinates $x
\mapsto x-a$ and deal with the system (\ref{1})--(\ref{3}) with
the shifted potential $q_a:=q(x-a)$, the objects corresponding to
$q_a$ being labelled with the subscript $a$. An $a \in {{\mathbb R}^3}$ is
said to be a {\it $s$-point of the potential $q$} if ${{\cal D}}_a
\not=\{0\}$ holds. By ${{\Upsilon_q}}$ we denote the set of such points.

To the best of our knowledge, $s$-points is something new in the
acoustical scattering and the set ${{\Upsilon_q}}$ is worth studying for its
own sake. The goal of our paper is to originate such a study: we
reveal certain relations between ${{\Upsilon_q}}$ and the known objects of
the scattering theory.

\subsection{Results}
Evolution of the system (\ref{1})--(\ref{3}) is governed by the
Schr$\ddot{\rm o}$dinger operator $H:=-\Delta + q$ in ${{\cal H}}$, ${\rm
Dom\,} H=H^2({{\mathbb R}^3})$. This operator may have a finite discrete
spectrum ${\sigma_{\rm disc}(H)} \subset (-\infty, 0]$ (see, e.g., \cite{RS}) and
in the paper, for the sake of simplicity, we accept
\begin{convention}\label{c1}
The potential $q$ is such that the equation
\begin{equation}\label{convention}
\left(-\Delta + q \right)\varphi=0 \qquad {\rm in}\,\,{{\mathbb R}^3}
\end{equation}
has not nonzero solutions, which satisfy $\varphi(x) \to 0$ as
$|x|\to 0$.
\end{convention}
In physical terms, this means that the hamiltonian $H$ has neither
bound nor semi-bound state at the zero energy level (see, e.g.,
\cite{Newt_book}). If exists, such a level can be removed by
arbitrarily small perturbation of the potential $q$, so that the
convention is not restrictive and we deal with a generic case.

Our results are the following.
\begin{itemize}
\item Let $S_a(k), \,k \in \mathbb R$ be the Schr$\ddot{\rm
o}$dinger $S$-matrix of the potential $q_a$. In the series of
papers cited in his book \cite{Newt_book}, R.Newton introduced the
Riemann--Hilbert factorization of the form
\begin{equation}\label{fact}
S_a(k)=\Pi^-_a(k)\left[S^-_a(k) S^+_a(k)\right]
\Pi^+_a(k)\,,\,\,\,\,k \in \mathbb R\,, \end{equation} where
$\Pi^{\pm}_a$ are the Blaschke-type rational operator-valued
functions, $S^\pm_a$ are the operator-valued functions holomorphic
and boundedly invertible in the half-planes $\{k \in {\mathbb C}\,
|\,\pm \Im k>0\}$ respectively. This representation is used for
solving the inverse problem that is to determine the potential $q$
from the scattering data: it is shown that the knowledge of the
family $\left\{S^\pm_a\right\}_{a \in {{\mathbb R}^3}}$ enables
one to recover $q$. In the mean time, in the mentioned papers it
is assumed that such a factorization is realizable for any $a \in
{{\mathbb R}^3}$. {\it We show that for $a \in {{\Upsilon_q}}$ the
representation (\ref{fact}) fails, i.e., the factorization is
impossible} \footnote{However, we do not claim that this fact
cancels the determination of the potential by R.Newton's
procedure.}. \item In \cite{BV3} we suggested that the presence of
$s$-points is connected with the discrete spectrum of the operator
$H$: the conjecture was that ${{\Upsilon_q}} \not=\emptyset$ is
equivalent to ${\sigma_{\rm disc}(H)}\not=\emptyset$. Here this
conjecture is partially justified as follows.

Let a point $a \in {{\mathbb R}^3}$, an integer $m\geq 1$, and a function $h
\in {{\cal H}}$ be such that
\begin{equation}\label{(-Delta+q)^j=0}
\left(-\Delta + q \right)^m h = 0\qquad {\rm in}\,\, {{\mathbb R}^3}
\setminus\{a\} \end{equation} holds, whereas ${\left(-\Delta+q\right)}^{m-1}h$ does not
vanish identically. As is shown in \cite{BV3}, such an $a$ is
necessarily an $s$-point; by ${{\Upsilon_q}}^m \subset {{\Upsilon_q}}$ we denote the set
of these points and specify $m$ as the order of $a$. {\it We prove
that ${{\Upsilon_q}}^1 \not=\emptyset$ is equivalent to
${\sigma_{\rm disc}(H)}\not=\emptyset$}. \item We show that the points $a \in
{{\Upsilon_q}^{\rm fin}}:=\cup_{m\geq 1}{{\Upsilon_q}}^m$  \footnote{Presumably, ${{\Upsilon_q}^{\rm fin}}$ exhausts
${{\Upsilon_q}}$ but it is still a conjecture.} can be specified as the {\it
jet degeneration points of polynomially growing solutions} to the
equations $\left(-\Delta + q \right) p = 0$ in ${{\mathbb R}^3}$. By this,
typically the set ${{\Upsilon_q}^{\rm fin}}$ consists of $s$-surfaces. If the
potential is radially symmetric, i.e., $q=q(|x|)$, the
$s$-surfaces are spheres and their position in the space can be
studied in more detail.
\end{itemize}

The work is supported by the RFBR grants 08-01-00511 and
NSh-4210.2010.1.

\section{$s$-points and R.Newton's factorization}
\subsection{Scattering matrix}
Recall the basic definitions and facts (see, e.g.,
\cite{Newt_book}).
\smallskip

The Schr$\ddot{\rm o}$dinger {\it scattering operator} of the pair
$H_0=-\Delta,\,\,H=-\Delta+q$ is $S: {{\cal H}} \to {{\cal H}}, \,\,S:=W^\ast_+
W_-$, where $W_\pm:={\rm s-lim}_{t \to \pm
\infty}e^{-itH}e^{itH_0}$ are the {\it wave operators}.

Let $\left(F y \right)(p)=\left(2\pi\right)^{-{3 \over
2}}\int_{{{\mathbb R}^3}}e^{i p x}y(x)\,dx$ be the Fourier transform. The
ope\-rator $\tilde S:=F S F^{-1}$ is of the form
$$\left(\widetilde S v \right)(p)=\left(S(k) \left[v(k,\,\cdot\,)\right]\right)(\theta)\,,$$
where $v(k,
\omega):=v(k\omega),\,\,k:=|p|,\,\,\theta:=\frac{p}{|p|},\,\,\omega,
\theta \in S^2:=\{p \in {{\mathbb R}^3} |\,|p|=1\}$, whereas $S(k)$ is an
operator acting in $L_2(S^2)$ on the angle variable by the rule
$$\left( S(k)\,g\right)(\theta) := \int_{S^2} s(\theta,
\omega; k) g(\omega)\,d\omega\,,$$ depending on $k$ as a parameter
and being called the {\it S-matrix} of the potential $q$. So, we
regard the S-matrix as an operator-valued function of $k>0$, the
values being taken in the bounded operator algebra ${\cal
B}\left(L_2(S^2)\right)$. The kernel $s(\theta, \omega; k)$ is a
distribution of the form $\delta(\theta-\omega)+ \widetilde
s(\theta, \omega; k)$ with a smooth $\widetilde s$, so that each
$S(k)$ is of the form "identity + compact operator". Moreover,
each $S(k)$ is a unitary operator. Also, the well-known high
energy asymptotic $S(k) \to \mathbb I$ as $k \to \infty$ holds.

In what follows, we assume $S(k)$ extended to $k<0$ by
$S(k):=S^\ast(-k)$. The {\rm index} of the S-matrix is defined by
$${\rm ind} S := \frac{1}{4\pi i}\left[\ln\det S(k)\right]\bigr|^{k=+\infty}_{k=-\infty}\,;$$
by the Levinson theorem it is equal to the total multiplicity of
the discrete spectrum ${\sigma_{\rm disc}(H)}$. The representation
\begin{equation}\label{Sred}
S(k)=\Pi^-(k)\,{S^{\rm red}}(k)\,\Pi^+(k)\,,\qquad k \in \mathbb R
\end{equation}
is valid, where ${S^{\rm red}}$ is the so-called {\it reduced S-matrix},
which satisfies ${\rm ind} {S^{\rm red}}=0$, and $\Pi$ is a ${\cal
B}\left(L_2(S^2)\right)$-valued function of the form
$$\Pi^\pm(k)=\prod \limits_{n=1}^{l}\left({\mathbb I}+\frac{2ik}{k-ik_n}\,B^\pm_n \right)\,, \qquad k \in \mathbb R$$
with the certain projections $B^\pm_n=\left(\,\cdot\,, \psi^\pm_n
\right)_{L_2(S^2)}\psi^\pm_n$, where
$\psi^+(\omega)=\psi^-(-\omega)$, and $k_n>0$ such that $-k^2_n\in
{\sigma_{\rm disc}(H)}$ (see \cite{Newt_book}).

\subsection{Factorization by R.Newton}
In solving the scattering inverse problem by R.Newton, the
following Riemann-Hilbert type representation of the reduced
S-matrix plays the key role:
\begin{equation}\label{factSr}
{S^{\rm red}}(k)=S^-(k) S^+(k)\,, \qquad k \in \mathbb R\,,
\end{equation}
where $S^\pm$ are the ${\cal B}\left(L_2(S^2)\right)$-valued
operator-functions, which are holomorphic, bounded, and boundedly
invertible in the complex half-planes ${\mathbb C}^\pm=\{k \in
{\mathbb C}\,|\, \pm \Im k >0\}$ respectively. In more detail, to
recover the potential $q$ via its S-matrix, one needs
\begin{itemize}
\item for a fixed $a \in {{\mathbb R}^3}$, to find the S-matrix $S_a$ of the
shifted potential $q_a$
 that can be done in a simple and
explicit way \item to determine ${S^{\rm red}}_a$ from (\ref{Sred}) and
represent ${S^{\rm red}}_a=S^-_a S^+_a$ by (\ref{factSr}) \item by varying
$a$, to collect the families $\left\{S^\pm_a\right\}_{a \in {{\mathbb R}^3}}$;
each of them determines the potential $q$.
\end{itemize}
However, in \cite{Newt_book} the author assumes that the second
step can be fulfilled {\it for  all $a \in {{\mathbb R}^3}$}. The
following result shows that such an assumption may be invalid.
\begin{theorem}\label{l1}
If $a \in {{\Upsilon_q}}$ then the representation (\ref{factSr}) for ${S^{\rm red}}_a$
does not hold, i.e., the factorization ${S^{\rm red}}_a=S^-_a S^+_a$ is
impossible.
\end{theorem}
We postpone the proof till section 1.4 and begin with
preliminaries concerning to the well-known facts of the
Lax-Phillips theory.

\subsection{On Lax-Phillips scheme}
The Cauchy problem for the acoustical equation is the system
\begin{align}
\label{Cuachy1} &v_{tt}-\Delta v+qv=0, \qquad (x,t) \in {{{\mathbb R}^3}} \times (-\infty,\infty) \\
\label{Cauchy2} &v \mid_{t=0} =\varphi, \quad v_t \mid_{t=0} =\psi
\qquad {\rm in}\,\,\,{{\mathbb R}^3}
\end{align}
with the finite energy data $d:=\{\varphi, \psi\}\in
D:=H^1({{\mathbb R}^3})\times L_2({{\mathbb R}^3})$; the solution is denoted by
$v=v^d(x,t)$. A peculiarity of the acoustical scattering is that
the {\it energy form} $$E[d,d']:=\int_{{{\mathbb R}^3}} \psi \psi'+\nabla
\varphi \cdot \nabla \varphi'+q \varphi \varphi'\,,$$ which the
set of data is equipped with, is indefinite. However, the form
turns out to be positive definite on the {\it absolutely
continuous subspace} $D_{\rm ac}:=D\ominus_E D_{\rm disc}$, where
$D_{\rm disc}$ is the finite dimensional subspace spanned on the
data $\{\varphi_k, \pm k \varphi_k\}$ such that $-k^2 \in {\sigma_{\rm disc}(H)}$
and $H\varphi_k=-k^2\varphi_k$. By $P_{\rm ac}$ we denote the
projection on the first summand of the decomposition $D=D_{\rm
ac}\oplus_E D_{\rm disc}$.

In the framework of the Lax-Phillips theory \cite{LP}, there is a
certain freedom in the choice of the pair of the {\it incoming}
and {\it outgoing subspaces} $D^\pm$. Once such a choice made, for
each $d \in D$ the corresponding {\it incoming} and {\it outgoing
spectral representatives} $\widetilde d_\mp$, which are the
$S^2$-valued functions of $k \in \mathbb R$ of the class
$L_2\left({\mathbb R}; L_2(S^2)\right)$, do appear. In the
spectral representation, the scattering process is described by
the {\it Lax-Phillips scattering operator} ${\widetilde S}: \widetilde d_-
\mapsto \widetilde d_+$, which acts in $L_2\left({\mathbb R};
L_2(S^2)\right)$ by
$$\left({\widetilde S} g\right)(k)\,=\,{\widetilde S}(k) g(k)\,, \qquad k \in \mathbb
R\,,$$ where ${\widetilde S} (\,\cdot\,)$ is a ${\cal B}\left(
L_2(S^2)\right)$-valued operator-function ({\it the Lax-Phillips
$S$-matrix}). Two possible variants of the choice are the
following.
\begin{enumerate}
\item Assign a $d\in D$ to the subspace $D^-_0\subset D$ if $v^d
\mid_{|x|<-t}=0$ for $t<0$ and the subspace $D^+_0\subset D$ if
$v^d \mid_{|x|<t}=0$ for $t>0$. The subspaces $$D^\mp\, :=\,P_{\rm
ac}D^\mp_0$$ constitute an incoming/outgoing pair \cite{LP}.  By
${\widetilde S}$ and ${\widetilde S}(\,\cdot\,)$ we denote the corresponding
scattering operator and its $S$-matrix. \item One can reduce the
incoming/outgoing subspaces to the smaller ones $$D^\mp_{\rm
red}\,:=\,D^\mp_0\cap D_{\rm ac} \subset D^\mp.$$  By ${{\widetilde S}^{\rm red}}$ and
${{\widetilde S}^{\rm red}}(\,\cdot\,)$ we denote the corresponding operator and
$S$-matrix.
\end{enumerate}

The important fact, which relates the quantum scattering and
acoustical scattering objects and follows from the above accepted
definitions, is that the equalities
\begin{equation}\label{S=S}
{\widetilde S}(k)\,=\,S(k)\,, \quad {{\widetilde S}^{\rm red}}(k)={S^{\rm red}}(k)\,, \qquad \,\,\,k \in
\mathbb R
\end{equation}
hold (see \cite{LP}, chapter VI, part 2, the relation (3.4)). Note
in addition that (\ref{S=S}) and (\ref{Sred}) imply
\begin{equation*}
{\widetilde S}(k)=\Pi^-(k)\,{{\widetilde S}^{\rm
red}}(k)\,\Pi^+(k)\,,\qquad k \in \mathbb R\,,
\end{equation*}
whereas the factors $\Pi^\mp(\,\cdot\,)$ can be also interpreted
in terms of the Lax-Phillips theory as $S$-matrices corresponding
to the certain choice of the incoming/outgoing subspaces.

Let \begin{equation}\label{Hardy} L_2\left({\mathbb R};
L_2(S^2)\right)\,=\,{{\cal H}}^- \oplus {{\cal H}}^+
\end{equation}
be the decomposition on the Hardy subspaces ${{\cal H}}^\mp$ which consist
of functions holomorphic in $\{k \in {\mathbb C}\, |\,\mp \Im
k>0\}$ respectively; by $P^\mp$ we denote the projections on
${{\cal H}}^\mp$. As is known, if $d$ belongs to the incoming (outgoing)
subspace then its spectral representative lies in ${{\cal H}}^-$ (${{\cal H}}^+$).
Representing the scattering operator in the matrix form in
accordance with (\ref{Hardy}), one has
\begin{equation}\label{blocks}
{{\widetilde S}^{\rm red}} =
\begin{pmatrix}
      P^- {{\widetilde S}^{\rm red}} P^-
         & P^- {{\widetilde S}^{\rm red}} P^+   \\
   {{\cal P}}^+ {{\widetilde S}^{\rm red}}   P^-
& P^+ {{\widetilde S}^{\rm red}} P^+ \end{pmatrix}\,.
\end{equation}

\subsection{Proof of Theorem \ref{l1}}
Assume that $a \in {{\mathbb R}^3}$ is a $s$-point of the potential $q$. As is
evident, by shifting the coordinate system, one can provide $a=0$.
So, let $0 \in {{\Upsilon_q}}$.

Let $w^f$ be a reversing wave (see section 0.2),
$d_0:=\{w^f(\,\cdot\,,0), 0\}$ its Cauchy data. By (\ref{cones}),
one has $d_0 \in D^-\cap D^+$; moreover, the results \cite{BV3} on
the stability of trajectories easily imply $d_0 \in D^-_{\rm
red}\cap D^+_{\rm red}$. The latter leads to $\widetilde d^-_0 \in
{{\cal H}}^-$ and $\widetilde d^+_0 \in {{\cal H}}^+$. Therefore,
\begin{equation*} {{\widetilde S}^{\rm red}} \widetilde d^-_0\,=\,\widetilde d^+_0\,\in {{\cal H}}^+\end{equation*}
and, applying the projection $P^-$, one gets $P^-{{\widetilde S}^{\rm red}} \widetilde
d^-_0 =P^-{{\widetilde S}^{\rm red}} P^- \widetilde d^-_0=0$. Thus, we see that ${\rm
Ker\,}P^-{{\widetilde S}^{\rm red}} P^- \not=\{0\}$, i.e., the block $P^-{{\widetilde S}^{\rm red}} P^-$ in
(\ref{blocks}) is not invertible.

By (\ref{S=S}), the block $P^-{S^{\rm red}} P^-$ is also not
invertible. However, as is well known, such an invertibility is
the necessary condition that provides the Riemann-Hilbert
factorization (\ref{factSr}) (see, e.g., \cite{Newt_book}). Hence,
in the case of $a \in {{\Upsilon_q}}$, to represent ${S^{\rm
red}}_a$ in the form (\ref{factSr}) is impossible. $\square$
\medskip

In addition, note the following. Although the representation
(\ref{factSr}) fails for $a \in {{\Upsilon_q}}$, by very general results of
the Riemann-Hilbert theory one can factorize the reduced S-matrix
as
\begin{equation}\label{factSred}
{S^{\rm red}}_a(k)=\check \Pi^-_a(k)\left[\check S^-_a(k) \check
S^+_a(k)\right]\check \Pi^+_a(k)\,, \qquad k \in \mathbb R\,
\end{equation}
with the Blaschke-type operator functions $\check \Pi_a, \,\check
\Pi^*_a$, which have the certain complex poles, and $\check
S^{\pm}_a$ holomorphic and boundedly invertible in $\{k \in
{\mathbb C}\, |\,\pm \Im k>0\}$ \cite{GoKrein}. Moreover, it is
the presence of the Blaschke factors, which renders the
representation (\ref{factSr}) impossible. In this connection, the
intriguing questions arises: What is the physical meaning of the
factorization (\ref{factSred}) and the corresponding Blaschke
poles? Can one characterize them in terms of the Lax-Phillips
theory? The questions are open.

\section{$s$-points and discrete spectrum}
\subsection{Theorem 2}
Assume that for a point $a\in {{\mathbb R}^3}$ there exist a function $h \in
{{\cal H}}$ and an integer $k\geq 1$ such that
$$\left(-\Delta + q \right)^k h = 0\qquad {\rm in}\,\,
{{\mathbb R}^3} \setminus\{a\}$$ and let $m$ be the minimal of such $k$'s. As
is shown in \cite{BV3}, this $a$ is necessarily an $s$-point,
whereas $m$ is specified as the order of $a$. By ${{\Upsilon_q}}^m \subset
{{\Upsilon_q}}$ we denote the set of $s$-points of the order $m$.
\smallskip

\noindent{\bf Remark}\,\,\,In the light of these definitions,
Convention \ref{c1} is motivated as follows. If $0 \in \sigma_{\rm
disc}(H)$ and $\varphi_0 \in {{\cal H}}$ is a zero energy level
eigenfunction, then $\left(-\Delta + q \right) \varphi_0 = 0$ {\it
everywhere} in ${{\mathbb R}^3}$ and we are forced to accept ${{\Upsilon_q}}^1={{\Upsilon_q}}={{\mathbb R}^3}$.
The convention excludes such a degeneration.
\smallskip

The following result partially explains why $s$-points do appear:
one of the reasons is the presence of the discrete spectrum of the
operator $H$.
\begin{theorem}\label{l2}
${{\Upsilon_q}}^1\not=\emptyset$ holds if and only if $\sigma_{\rm
disc}(H)\not=\emptyset$.
\end{theorem}
The rest of section 2 is the proof of the theorem.

\subsection{Green function} We define the {\it Green function} of the
operator $H$ as a $L_2^{\rm loc}({{\mathbb R}^3})$ -solution of the integral
equation
\begin{equation}\label{GreenFEq}
G(x,y)= \frac{1}{4\pi|x-y|} -
\int_{{{\mathbb R}^3}}\frac{q(s)}{4\pi|x-s|}\,G(s,y)\,ds\,,\qquad x
\,\,\in\,{{\mathbb R}^3}\,,
\end{equation}
where $y \in {{\mathbb R}^3}$ is a parameter. This definition is correct since
this equation is uniquely solvable. Indeed, otherwise there is a
nonzero solution $\varphi$ of the homogeneous equation
\begin{equation*}
\varphi(x)=\,-\,
\int_{{{\mathbb R}^3}}\frac{\varphi(s)}{4\pi|x-s|}\,q(s)\,ds\,,\qquad
x\,\,\in\,{{\mathbb R}^3}\,.
\end{equation*}
As is easy to see, such a solution satisfies (\ref{convention})
and vanishes at infinity that is forbidden by Convention \ref{c1}.
To establish the solvability, one can reduce the equation to a
large enough ball $B\supset {\rm supp\,} q$, prove the solvability
there, and then extend the solution to ${{\mathbb R}^3} \backslash B$ as a
r.h.s. of (\ref{GreenFEq}).

Also, the integral equation implies the symmetry property $G(x,
y)=G(y, x),\,\,\,x\not= y$. From (\ref{GreenFEq}) one derives that
in the sense of distributions on $C^\infty_0({{\mathbb R}^3})$ the function
$G$ satisfies
\begin{equation}\label{HG=delta}
\left(-\Delta+q \right)G(\,\cdot\,,y) = \delta_y\,,
\end{equation}
where $\delta_y$ is the Dirac measure supported in $y$.
\begin{lemma}\label{G=Phi/|x|+O} The asymptotic
\begin{equation}\label{asymptoticG}
G(x,y) \underset{|x| \to \infty}=
\frac{\Phi(y)}{4\pi|x|}\,+\,O\left(\frac{1}{|x|^2}\right)\,,
\qquad y \in {{\mathbb R}^3}
\end{equation}
holds, where $\Phi$ is a smooth function obeying
\begin{align}
\label{EqnPhi}& \left(-\Delta+q \right)\Phi = 0 \qquad {\rm
in\,}\,\, {{\mathbb R}^3}\\
\label{asymptoticPhi}& \Phi(x) \to 1 \quad \,{\rm as} \,\,\,|x|
\to \infty\,,
\end{align}
the asymptotics (\ref{asymptoticG}) and (\ref{asymptoticPhi})
being uniform w.r.t. $\frac{x}{|x|}\in S^2$.
\end{lemma}
{\bf Proof}\,\,\,Fix $x$ in the equation (\ref{GreenFEq}) and tend
$|y|\to \infty$. Using the symmetry property and looking for the
solution in the form (\ref{asymptoticG}), we have
\begin{equation*}
\frac{\Phi(x)}{4\pi|y|} + O\left(\frac{1}{|y|^2}\right)=
\frac{1}{4\pi|y|} + O\left(\frac{1}{|y|^2}\right)-
\int_{{{\mathbb R}^3}}\frac{q(s)}{4\pi|x-s|}\,
\left[\frac{\Phi(s)}{4\pi|y|}+O\left(\frac{1}{|y|^2}\right)\right]\,ds
\end{equation*}
that easily leads to the integral equation
\begin{equation}\label{IntEqnPhi}
\Phi(x)= 1 -
\int_{{{\mathbb R}^3}}\frac{q(s)}{4\pi|x-s|}\,\Phi(s)\,ds\,,\qquad x
\,\,\in\,{{\mathbb R}^3}\,.
\end{equation}
The latter equation implies (\ref{EqnPhi}), (\ref{asymptoticPhi}),
and justifies (\ref{asymptoticG}). \,\,\,$\square$
\smallskip

Now we are ready for proving Theorem \ref{l2}.

\subsection{Necessity}
Let $\sigma_{\rm disc}(H)\not=\emptyset$ and $H\chi=-k^2_0 \chi$
for $- k^2_0= \inf \sigma_{\rm disc}(H)$, so that $\chi$ is a
ground state of the operator $H$. As is well known, $-k^2_0$ is an
ordinary eigenvalue of $H$, whereas the eigenfunction $\chi$
behaves as $\chi(x) \underset{|x|\to \infty}\sim e^{-k_0|x|}$ and
can be chosen positive: $\chi>0$ {\it everywhere} in ${{\mathbb
R}^3}$. Integrating by parts, we have
\begin{equation*}
0=\langle {\rm see\,\,(\ref{EqnPhi})} \rangle =\int_{{{\mathbb R}^3}}\chi(x)
\left[\left(-\Delta+q \right)\Phi \right](x)\,dx= -k^2_0
\int_{{{\mathbb R}^3}}\chi(x) \Phi(x)\,dx\,.
\end{equation*}
Hence, $\Phi$ has to change sign and there is an $a \in {{\mathbb R}^3}$ such
that $\Phi(a)=0$. As result, we conclude that $G(\,\cdot\,,a) \in
{{\cal H}}$ (see (\ref{asymptoticG})), whereas $\left(-\Delta+q
\right)G(\,\cdot\,,a)=0\,\,{\rm in}\,\,{{\mathbb R}^3} \setminus\{a\}$ (see
(\ref{HG=delta})). Thus, $a \in {{\Upsilon_q}}^1$ and, hence,
${{\Upsilon_q}}^1\not=\emptyset$.

\subsection{Sufficiency}
The sufficiency will be proved by contradiction. Assume that
$\sigma_{\rm disc}(H)=\emptyset$ but ${{\Upsilon_q}}^1\not=\emptyset$ and,
hence, there is a point $a \in {{\mathbb R}^3}$ and a function $h \in {{\cal H}}$ such
that $\left(-\Delta+q \right)h=0$ holds in ${{\mathbb R}^3} \setminus \{a\}$.

Show that $\Phi(a)=0$. Indeed, considering $\left(-\Delta+q
\right)h$ as a distribution on $C^\infty_0\left({{\mathbb R}^3}\right)$, we
see that it is supported at $x=a$. Such a distribution has to be a
linear combination of the Dirac measure derivatives:
$$\left(-\Delta+q \right)h=\sum \limits^N_{|j|=0}\alpha_j D^j_x \delta_a=
\sum \limits^N_{|j|=0}\alpha_j (-1)^{|j|}D^j_a \delta_a\,,$$ where
$j=\{j_1, j_2, j_3\}$ is a multi-index, $|j|=j_1+j_2+j_3$ and
$D^j_x=\partial_{x^1}^{j_1} \partial_{x^2}^{j_2}
\partial_{x^3}^{j_3}$ is a differentiation, $\alpha_j$ are
constants (see, e.g, \cite{V}). Therefore, by (\ref{HG=delta}),
the function $h$ has to be of the form
\begin{equation}\label{h=G+tildeh}h(x)= \sum
\limits^N_{|j|=0}\alpha_j(-1)^{|j|} D^j_a G(x, a)+\widetilde
h(x)\,,\end{equation} with a $\widetilde h$ satisfying
$-(\Delta+q)\widetilde h=0$ as a distribution. Hence, $\widetilde
h$ is a smooth function and
\begin{equation}\label{auxill_1}
-(\Delta+q)\widetilde h=0 \qquad {\rm in}\,\,{{\mathbb R}^3}
\end{equation}
holds in the classical sense. As can be shown from
(\ref{GreenFEq}), for $|j|\geq 1$ the derivatives $D^j_a
G(\,\cdot\,, a)$ are not square-summable near $x=a$. Therefore,
the only option for the sum in (\ref{h=G+tildeh}) (and, hence, for
$h$) to belong to ${{\cal H}}$ (to be square-summable near $x=a$) is
$\alpha_j=0$ for $|j|\geq 1$ and we get
$$h(x)={\alpha_0}G(x, a)+\widetilde h(x)\, \qquad
{\rm in}\,\,{{\mathbb R}^3}\backslash\{0\}.$$ For large enough $|x|>{\rm
diam~supp\,}q$, we have $\Delta h =0$ and know that $h \in {{\cal H}}$.
Such a function vanishes as $|x| \to \infty$. Hence, by
(\ref{asymptoticG}), the function $\widetilde h$ also tends to
zero as $|x| \to \infty$ and, in the same time, satisfies
(\ref{auxill_1}). By Convention 1,  one has $\widetilde h=0$.
Thus, $h$ is proportional to the Green function $G(\,\cdot\,,a)$,
whereas $G(\,\cdot\,,a)\in {{\cal H}}$ implies $\Phi(a)=0$ by
(\ref{asymptoticG}).

Now we apply the perturbation theory arguments. Consider an
operator family $\{-\Delta+\varepsilon q\}_{\varepsilon \in
[0,1]}$; let ${\Phi^\varepsilon}$ be the corresponding analog of the function
$\Phi \equiv \Phi^1$ that appears in the asymptotic
(\ref{asymptoticG}). Note that for $\varepsilon=0$ one has
$\Phi^0=1$. As is easy to see from the integral equation
(\ref{GreenFEq}), for small $\varepsilon$ the inequality
${\Phi^\varepsilon}(\,\cdot\,)>0$ holds everywhere in ${{\mathbb R}^3}$, whereas
\begin{equation}\label{Pepsto1}
{\Phi^\varepsilon}(x)\to 1 \quad {\rm as}\,\,\,|x| \to \infty
\end{equation}
uniformly w.r.t. $\varepsilon$.

Since ${\Phi^\varepsilon}$ depends on $\varepsilon$ continuously \footnote{in
fact, analytically}, there is the minimal $\varepsilon_0 \in
(0,1]$ such that the function ${\Phi^{\varepsilon_0}}$ satisfies ${\Phi^{\varepsilon_0}} \geq 0$
and, in the mean time, has zeros. Shifting (if necessary) the
origin of coordinates, assume that ${\Phi^{\varepsilon_0}}(0)=0$ and recall that
${\Phi^{\varepsilon_0}}$ satisfies $\left(-\Delta+\varepsilon_0 q \right){\Phi^{\varepsilon_0}}=0$
in ${{\mathbb R}^3}$.

Fix a positive $r$; let $B_r:=\{x \in {{\mathbb R}^3}\,|\,|x|< r\}$ and
$S_r:=\partial B_r$. Integration by parts implies
\begin{align*}
& \varepsilon_0 \int_{B_r} q(\xi) {\Phi^{\varepsilon_0}}(\xi) \left({1 \over
{|\xi|}} -{1 \over r} \right)\,d\xi\,=\,\int_{B_r} \Delta
{\Phi^{\varepsilon_0}}(\xi)
\left({1 \over {|\xi|}} -{1 \over r} \right)\,d\xi\,=\\
& -4\pi {\Phi^{\varepsilon_0}}(0) + {1 \over {r^2}}\int_{S_r}
{\Phi^{\varepsilon_0}}(\omega)\,d\omega =\int_{S_1} {\Phi^{\varepsilon_0}}(r\theta)\,d\theta\,=:
U(r)\,.
\end{align*}
From the other hand, we have
\begin{align*}
& \biggl| \int_{B_r} q(\xi) {\Phi^{\varepsilon_0}}(\xi) \left({1 \over {|\xi|}}
-{1 \over r} \right)\,d\xi\,\biggr |\,\leq\,\underset{B_r} \max
\,|q| \int \limits^r_0 d\tau\,\tau^2 \left({1 \over \tau}-{1 \over
r} \right)\int_{S_1} {\Phi^{\varepsilon_0}}(\tau \theta)\,d\theta\,\\
& \leq\,\underset{B_r} \max \,|q|\,r\,\int \limits^r_0
U(\tau)\,d\tau \leq \left[\underset{B_r} \max\, |q| \right]\,
r^2\,\left[\underset{[0,r]}\max\, U \right]
\end{align*}
and arrive at the estimate
\begin{equation*} U(r)\,\leq\,c
r^2 \,\underset{[0,r]}\max\, U (\,\cdot\,)\,.
\end{equation*}
Choosing $r$ small enough such that $U(r)=\underset{[0,r]}\max\, U
(\,\cdot\,)$ we see that this estimate is possible only if
$U(r_1)$ vanishes identically for  $r_1  < r$. Since ${\Phi^{\varepsilon_0}} \geq
0$, this means that ${\Phi^{\varepsilon_0}}  \equiv  0$ in a small ball $B_r$.

So, ${\Phi^{\varepsilon_0}}$ is a solution of an elliptic equation vanishing in a
ball. By the Landis uniqueness theorem \cite{L}, such a solution
vanishes everywhere and we have ${\Phi^{\varepsilon_0}}=0$ in ${{\mathbb R}^3}$ that
contradicts to (\ref{Pepsto1}). $\square$

\section{$s$-points and degeneration of jets}
\subsection{Jets and polynomials} Recall what a {\it jet} is. For
a fixed $a \in {{\mathbb R}^3}$ and an integer $k\geq 0$, we say the smooth
functions $u$ and $v$ to be equivalent (and write $u\sim v$) if
$u(x)-v(x)=o\left( |x-a|^k\right)$ as $x \to a$. With respect to
the relation $\sim$, the set $C^\infty({{\mathbb R}^3})$  decays on the
equivalence classes, whereas the class $j^k_a[u]:=\{v\,|\,v \sim
u\}$ is called the jet of $u$ (at the point $a$, of the order
$k$). Setting up $\alpha j^k_a[u]+\beta j^k_a[u]:= j^k_a[\alpha
u+\beta v]$ for $\alpha, \beta \in \mathbb R$, one makes the set
of jets into a linear space ${{\cal J}}_a^k\left[C^\infty({{\mathbb R}^3}) \right]$.

As is easy to recognize, the jet $j^k_a[u]$ can be identified with
the collection of the derivatives $\{D^j_x u(a)\}_{|j|=0}^k$,
whereas ${\rm dim\,}{\cal J}_a^k\left[C^\infty({{\mathbb R}^3}) \right]$ is
equal to the number $d^k$ of such derivatives: $d^0=1,\,
d^1=1+3=4,\,d^2=1+3+6=10,\,\dots\,,
d^k=\frac{1}{6}(k+1)(k+2)(k+3)\,$.

Let ${\cal L} \subset C^\infty({{\mathbb R}^3})$ be a linear set; introduce a
subspace
$${{\cal J}}_a^k\left[{\cal L}\right]:=\left\{j^k_a[u]\,|\,\,u \in {\cal
L} \right\} \subset {{\cal J}}_a^k\left[C^\infty({{\mathbb R}^3}) \right]$$ and denote
by $d^k_a[\cal L]$ its dimension.

Fix an integer $l\geq 0$; the functions belonging to a linear set
$$
{{\cal P}}^l_q:=\left\{p \in C^\infty({{\mathbb R}^3})\,|\,\left(-\Delta+q
\right)p=0\,\,{\rm in}\,{{\mathbb R}^3},\,\,|p(x)|\leq {\rm
const}\left(1+|x|\right)^l\right\}\,.
$$
are said to be {\it $q$-harmonic polynomials} of the order $\leq
l$. The existence of such functions is established in a standard
way: one puts $p=p_l+w$, where $p_l$ is a harmonic polynomial
(see, e.g., \cite{V}), derives the relevant integral equation for
$w$ of the form analogous to (\ref{GreenFEq}), and proves its
solvability and uniqueness of the solution by the same arguments
as for the equation (\ref{GreenFEq}), i.e., referring to
Convention \ref{c1}. For instance, the function $\Phi$ determined
by (\ref{EqnPhi}), (\ref{asymptoticPhi}) is an element of ${{\cal
P}}^0_q$. Also, as is easy to verify, ${{\cal P}}^{l}_q\subset
{{\cal P}}^{l'}_q$ for $l < l'$, and the relations
\begin{equation}\label{jetdim}{\rm dim}{{\cal P}}^l_q={\rm
dim}{{\cal P}}^l_0=(l+1)^2, \quad d^k_a[{{\cal P}}^k_q] \leq
d^k_a[{{\cal P}}^k_0]=(k+1)^2\end{equation} hold.

The main result of this section is
\begin{theorem}\label{l3}
For $m=2,3, \dots$, the inclusion $a \in {{\Upsilon_q}}^m$ is equivalent to
the relation
\begin{equation}\label{jetdiminequality}
d^{2m-2}_a[{{\cal P}}^{2m-2}_q] < (2m-1)^2\,.
\end{equation}
\end{theorem}
Since in the generic case one has $d^{2m-2}_a[{{\cal P}}^{2m-2}_q] =
(2m-1)^2$ (see (\ref{jetdim}) for $k=2m-2$), it is reasonable to
refer to (\ref{jetdiminequality}) as a jet degeneration of
${{\cal P}}^{2m-2}_q$ at the point $x=a$. Also, the Theorem \ref{l2} can
be interpreted in the same terms: $\Phi(a)=0$ means that ${\rm
dim\,}{{\cal J}}^0_a\left[{{\cal P}}^0_q\right]=0<1$.

The proof is postponed till section 3.3. For the sake of
simplicity, it will be demonstrated for the case $m=2$; the way to
treat the general case will be clear.

\subsection{Green function}
Here we deal with the $q$-biharmonic equation $\left(-\Delta+q
\right)^2 u=h$. For $x, y \in {{\mathbb R}^3}$, $\langle x,y\rangle:=x_1 y_1 +
x_2 y_2 + x_3 y_3$ is the inner product.
\smallskip

\noindent{\bf Unperturbed case}\,\,\,For $q=0$, we set
$$B_0(x,y)\,:=\,- \frac{|x-y|}{8\pi}$$ and, in view of the well-known
relation
\begin{equation}\label{Delta^2B0=delta}
\left(-\Delta \right)^2B_0(\,\cdot\,,y)= \delta_y\,,
\end{equation} refer to $B_0$ as the {\it
unperturbed Green function}. The following is a standard way to
derive the asymptotic of $B_0$ as $|x| \to \infty$:
\begin{align*}
&| x  -y|  =  \left( |x|^2  -2\langle x,y\rangle +|y|^2\right)^{\frac{ 1 }  { 2   }} =\\
&|x|    \left ( \,\,1   +\frac{ |y|^2 }  { |x|^2   }   -\frac{ 2 }
{ |x|   }   \left\langle \frac{ x }  { |x|   },y  \right \rangle
  \right  )   ^{\frac{ 1 }  { 2   }}=\\
&|x|    \left ( \,\,1   + \frac{ 1 }  { 2   }
 \left(
      \frac{ |y|^2 }  {|x|^2   }   -\frac{ 2 }  { |x|   }   \left\langle \frac{ x }  { |x|   },y  \right\rangle
\right ) -\frac{ 1 }  { 8   }
 \left(
      \frac{ |y|^2 }  { |x|^2   }   -\frac{ 2 }  { |x|   }   \left\langle \frac{ x }  { |x|   },y  \right
      \rangle
  \right )   ^ 2   +O( |x|  ^ {-3   })
  \right  )   =\\
&|x|    \left ( \,\,1   +
                  \frac{ |y|^2 }  {2 |x|^2   }   -\frac{ 1 }  { |x|   }
                  \left\langle \frac{ x }  { |x|   },y  \right\rangle
                             -\frac{ 1 }  {2 |x|^2   }
                                         \left\langle \frac{ x }  { |x|   },y    \right \rangle   ^ 2   +
                                         O ( |x|  ^ {-3   })
\right )  =\\
&|x|    +
                  \frac{ |y|^2 }  {2 |x|   }   -   \left\langle \frac{ x }  { |x|   },y  \right \rangle
                               -\frac{ 1 }  {2 |x|   }
                          \left\langle \frac{ x }  { |x|   },y
                \right \rangle  ^ 2   + O ( |x|  ^ {-2   })=\\
&|x|    +
                  \frac{ |y|^2 }  {2 |x|   }   -
    \frac{ 1 }{ |x|   }   \sum_{i}
  x_i  y_i
- \frac{ 1 }{ |x| ^3  }   \sum_{i\neq j}
  x_i x_j \, y_i y_j
-  \frac{ 1 }{2 |x| ^3  }  \sum_{i}
  x_i^2 \, y_i^2 +           \\
&     +O  \left(   \frac{1}{|x|^2}\right)\,,
\end{align*}
where $i$ and $j$ run over 1,2,3. To clarify the structure of this
expression, one can use the identity
\begin{align*}
  \sum_{i}
  x_i^2\,y_i^2=
\frac{ 1 }  { 2   }  (  x_1^2-x_2 ^2 )\,  (  y_1^2-y_2 ^2 )+
  \frac{ 1 }  { 6   }  (  x_1^2+x_2^2 -2x_3^2 )\,  (  y_1^2+y_2 ^2 -2y_3^2 )
+\frac{ 1 }  { 3   } |x|^2 \,|y|^2
\end{align*}
and write the result in the form
\begin{align}
\notag &| x -y| \,= \left\{|x| + \frac{ |y|^2 }  {3 |x|
}\right\}\, -\, \left\{ \sum_{i} \left[\frac{x_i}{|x|}\right]\,
y_i\right\}\,-\,\frac{ 1 }{ |x|}  \biggl \{\sum_{i\neq
j}\left[\frac{x_i x_j }{ |x| ^2}\right] \,
y_i y_j\,-\,\\
\notag &\frac{ 1 }{4} \left[\frac{x_1^2-x_2 ^2}{ |x| ^2}\right]\,
( y_1^2-y_2 ^2 )\,-\,
  \frac{1}{12}\left[\frac{x_1^2+x_2^2 -2x_3^2}{|x|^2}\right]
\, ( y_1^2+y_2 ^2 -2y_3^2 )\biggr\}\\
\label{asympt|x-y|} & + O\left( \frac{1}{|x|^2}\right).
\end{align}
As is seen now, this is an expansion over the spherical harmonics:
the terms into the first curly braces do not depend on angle
variables, the ones in the second and third braces are
proportional to $Y^m_1(\frac{x}{|x|})$ and $Y^m_2(\frac{x}{|x|})$
respectively\footnote{the harmonics are selected by the square
brackets}. One more peculiarity of such a representation is that
among the $y$-dependent coefficients
\begin{align}
\notag & \Phi^0_1=1\,,\,\,\Psi^0=|y|^2\,,\\
\notag & \Phi^0_2=y_1\,,\,\,\Phi^0_3=y_2\,,\,\,\Phi^0_4=y_3\,,\\
\label{Phi^0Psi^0}& \Phi^0_5=y_1 y_2\,,\,\,\Phi^0_6=y_2
y_3\,,\,\,\Phi^0_7=y_1 y_3\,,\,\,\Phi^0_8=y_1^2-y_2
^2\,,\,\,\Phi^0_9=y_1^2+y_2 ^2 -2y_3^2
\end{align}
there are nine {\it harmonic} functions
$\Phi^0_1,\,\dots\,,\Phi^0_9$ and one {\it biharmonic} $\Psi^0$.
\begin{remark}So, the leading part of the unperturbed Green function
asymptotic (\ref{asympt|x-y|}) is of the following structure:
\begin{itemize} \item the term harmonic w.r.t. $x$ has the coefficient
$\Psi^0$ biharmonic w.r.t. $y$ \item the other terms are
biharmonic w.r.t. $x$ and have the coefficients $\Phi^0_i$
harmonic w.r.t. $y$.
\end{itemize}
\end{remark}
Quite analogous picture will be observed in the perturbed case.
\medskip

\noindent{\bf Perturbed case}\,\,\,For $q\not=0$, we {\it define}
the (perturbed) Green function as a $L_\infty^{\rm loc}({{\mathbb R}^3})$
-solution of the equation
\begin{align}\label{DefB(x,y)}
B(x,y)=-\frac{|x-y|}{8\pi}\,+\,
   \int_{{{\mathbb R}^3}} \frac{|x-t|}{8\pi}\,L B(t,y)\,dt\,, \qquad x \in
   {{\mathbb R}^3},
\end{align}
where $y \in {{\mathbb R}^3}$ is a parameter and $L:=\left(-\Delta+q
\right)^2-\left(-\Delta\right)^2$ is a second-order differential
operator\footnote{in the integral, $L$ acts on the variable $t$}
such that ${\rm supp}\,Lu \subset {\rm supp\,}q \cap {\rm
supp\,}u$. The following result shows that this definition is
correct. \begin{lemma}\label{lemma-uniqueness} The equation
(\ref{DefB(x,y)}) is uniquely solvable. \end{lemma} {\bf
Proof}\,\,\,Let the homogeneous equation
\begin{align*}
    \psi (x) =
   \int_{{{\mathbb R}^3}}\frac{|x-t|}{8\pi}\, L \psi (t)\,dt
\end{align*}
have a nonzero solution $\psi$. Such a solution satisfies
\begin{align*}
 \left (-\Delta +q \right)^2\psi=0 \quad {\rm in}\,\,\,{{\mathbb R}^3}\,;
\end{align*}
by  (\ref{asympt|x-y|}), for large $|x|$ it behaves as $$\psi(x)
\underset{|x| \to \infty}=\alpha |x|+\sum \limits_{i=1}^3\beta_i
\frac{x_i}{|x|}+\widetilde \psi(x)\,,$$ where $\widetilde \psi
=O(\frac{1}{|x|})\to 0$.

Assume that $\psi$ is $q$-harmonic: ${\left(-\Delta+q\right)} \psi=0$ in ${{\mathbb R}^3}$. If so,
we have $\Delta \psi=0$ for large $|x|$ that implies
$\alpha=\beta_i=0$ and $\psi=\widetilde \psi$. Hence, $\psi$ is a
$q$-harmonic function vanishing as $|x| \to \infty$ that is
forbidden by Convention \ref{c1}.

Thus, $ \varphi:=  \left ( -\Delta +q \right )\psi$ is a nonzero
function, which satisfies $ \left (-\Delta +q \right)\varphi=0$
in ${{\mathbb R}^3}$ and
\begin{align*}
\varphi(x) \underset{|x| \to \infty}=-\alpha \Delta |x|-\sum
\limits_{i=1}^3\beta_i \Delta \frac{x_i}{|x|}+\Delta\widetilde
\psi(x)\,=\,O\left(\frac{1}{|x|}\right)\to 0
\end{align*}
that is impossible in view of Convention \ref{c1}. So, assuming
$\psi \not=0$ we arrive at a contradiction. \,\,\,$\square$
\medskip

As easily follows from (\ref{DefB(x,y)}), the relation
\begin{equation}\label{(Delta-q)^2B=delta}
\left(-\Delta+q \right)^2B(\,\cdot\,,y)= \delta_y
\end{equation}
analogous to (\ref{Delta^2B0=delta}) is valid. Also, the symmetry
property $B(x, y)=B(y, x)$ holds.
\smallskip

The way to derive the asymptotic of $B$ analogous to
(\ref{asympt|x-y|}) is also quite standard. Namely, we represent
\begin{equation}\label{asympB}
B(x,y)=-\frac{1}{8\pi}\,A(x,y)+O(|x|^{-2}) \end{equation} with an
ansatz of the form\footnote{compare with (\ref{asympt|x-y|})!}
\begin{align}
\notag &A(x,y) \,= \left\{|x|\Phi_1(y) + \frac{\Psi(y)}  {3 |x|
}\right\}\, -\\
\notag & \left\{ \left[\frac{x_1}{|x|}\right]
\Phi_2(y)+\left[\frac{x_2}{|x|}\right]
\Phi_3(y)+\left[\frac{x_3}{|x|}\right]\Phi_4(y)
\right\}\,-\\
\notag &\frac{ 1 }{ |x|} \biggl \{\left[\frac{x_1 x_2}{ |x|
^2}\right]\Phi_5(y)\,+\,\left[\frac{x_2 x_3}{ |x| ^2}\right]
\Phi_6(y)\,+\,\left[\frac{x_1 x_3}{ |x|
^2}\right] \Phi_7(y)\,+\\
\label{ansatzA} &\frac{ 1 }{4} \left[\frac{x_1^2-x_2 ^2}{ |x|
^2}\right]\Phi_8(y)\,-\,
  \frac{1}{12}\left[\frac{x_1^2+x_2^2 -2x_3^2}{|x|^2}\right]
\Phi_9(y)\biggr\}\,,
\end{align}
tend $|y| \to \infty$ in (\ref{DefB(x,y)}), using the symmetry
property substitute this representation to (\ref{DefB(x,y)}), and
by equaling the coefficients at the proper spherical harmonics,
eventually get the 'transport equations' of the form
\begin{align}
\label{eqnPhi}& \Phi_i(x) = \Phi^0_i(x)\,+\,
   \int_{{{\mathbb R}^3}} \frac{|x-t|}{8\pi}\,L \Phi_i(t)\,dt\,, \qquad i=1, \dots, 9\\
\label{eqnPsi}& \Psi(x) = \Psi^0(x)\,+\,
   \int_{{{\mathbb R}^3}} \frac{|x-t|}{8\pi}\,L \Psi(t)\,dt\,.
\end{align}
By the same arguments, which provide the unique solvability of
(\ref{DefB(x,y)}), these equations are also solvable uniquely.

As is easy to see from (\ref{eqnPhi}) and (\ref{eqnPsi}), the
solutions $\Psi$ and $\Phi_i$ are q-biharmonic. Moreover,
\begin{lemma}
The functions $\Phi_1,\, \dots ,\,\Phi_9$ are q-harmonic
polynomials of the order $\,\,\leq 2$ and constitute a basis in
${{\cal P}}^2_q$.
\end{lemma}
{\bf Proof}

\noindent {\it Step 1: $q$-harmonicity}\,\,\,Take a $f \in
      C^\infty_0({{\mathbb R}^3}),$
 denote ${\hat{H}}:={{ \left(-\Delta+q\right) }}$
\footnote{   So, $ {\hat{H}}  $ is
understood as a differential expression that can be applied to any
smooth enough function not necessarily belonging to ${{\cal H}}$.} and put
\begin{equation}\label{F}
F(x):=\int_{{{\mathbb R}^3}}B(x,y)\,  \hat{H} f(y)\,dy\,, \qquad x \in {{\mathbb R}^3}.
\end{equation} Applying $\hat{H}^2$ and taking into account (\ref{(Delta-q)^2B=delta}),
we get ${\hat{H}}^2 F={\hat{H}} f$ and, hence,
$$      {\hat{H}}   \left( {\hat{H}} F - f\right)=0 \qquad {\rm in}\,\,{{\mathbb R}^3}.$$
As is easily seen from (\ref{F}) and (\ref{asympB}),
(\ref{ansatzA}), the function ${\hat{H}} F$ vanishes as $|x| \to
\infty$. Hence, the same is valid for ${\hat{H}} F-f$. In the mean time,
the latter difference is annulated by ${\hat{H}}$ and, whence, it has to
vanish identically by Convention \ref{c1}. Thus, we arrive at ${\hat{H}}
F - f=0$, i e.,
\begin{equation}\label{HF=f}
\left(-\Delta + q(x)\right) \int_{{{\mathbb R}^3}}B(x,y)\,{\hat{H}}
f(y)\,dy\,=f(x)\,, \qquad x \in {{\mathbb R}^3}.
\end{equation}

Now, tending $|x| \to \infty$ and taking into account the
structure of the asymptotic (\ref{asympB}), (\ref{ansatzA}), it is
easy to see that the asymptotic of the l.h.s. contains certain
linearly independent harmonic (w.r.t. $x$) terms\footnote{for
large enough $|x|$, 'harmonic' and 'q-harmonic' is the same} with
the coefficients of the form $\int_{{{\mathbb R}^3}}\Phi_i(y)\,{\hat{H}} f(y)\,dy$.
In the mean time, the r.h.s. in the equality (\ref{HF=f}) is
compactly supported and, hence, the equality forces all harmonic
terms to vanish identically. This follows to
$$0=\int_{{{\mathbb R}^3}}\Phi_i(y)\,{\hat{H}} f(y)\,dy=\int_{{{\mathbb R}^3}}
{\hat{H}}
\Phi_i(y)\,f(y)\,dy$$ that is equivalent to ${\hat{H}} \Phi_i=0$ in
${{\mathbb R}^3}$ by arbitrariness of $f$. So, $\Phi_i$ are q-harmonic.
\smallskip

\noindent {\it Step 2: integral equations}\,\,\, The functions
$\Phi_i$ satisfy the integral equations, which are different from
and more informative than (\ref{eqnPhi}). These equations can be
derived as follows.

By (\ref{HG=delta}), for any $f\in C^\infty_0({{\mathbb R}^3})$ one has the
relation
$$\left(-\Delta + q(x)\right) \int_{{{\mathbb R}^3}}G(x,
y)\,f(y)\,dy\,=\,f(x)\,, \qquad x \in {{\mathbb R}^3}\,. $$ Comparing it with
(\ref{HF=f}) we get $$\int_{{{\mathbb R}^3}}G(x,
y)\,f(y)\,dy\,=\,\int_{{{\mathbb R}^3}}B(x,y)\,{\hat{H}} f(y)\,dy\,,$$ whereas
integration by parts easily implies
\begin{equation}\label{HB=G}
\left(-\Delta + q(y)\right) B(x,y)\,=\,G(x,y)\,, \qquad x,y \in
{{\mathbb R}^3}, \,\,\,x\not= y.
\end{equation}
In the unperturbed case, for the Green functions $B_0$ and $G_0$
we have just
\begin{equation}\label{H0B0=G0}
-\Delta
\left[-\frac{|x-y|}{8\pi}\right]\,=\,\frac{1}{4\pi|x-y|}\,, \qquad
x,y \in {{\mathbb R}^3}, \,\,\,x\not= y.
\end{equation}
These relations enable one to get the asymptotics for $G$ and
$G_0$ through (\ref{asympB}) and (\ref{asympt|x-y|}) respectively,
{\it the coefficients of the asymptotics being expressed through
$\Phi_i, \Psi$ and $\Phi_i^0, \Psi^0$}. Surely, the asymptotic of
$G$ obtained in this way is just a more detailed version of
(\ref{asymptoticG}), which takes into account the structure of the
lower order terms. Also, the comparison of the coefficients at
$\frac{1}{|x|}$ in the detailed asymptotic and (\ref{asymptoticG})
easily implies the important relation
\begin{equation}\label{Phi=HPsi}
\Phi(y) \,=\,-\frac{1}{6}\left(\Delta + q(y)\right)\Psi(y)\,,
\end{equation}
which will be required later on.

Thereafter, the trick, which was used for derivation of the
equation (\ref{IntEqnPhi}), is repeated: substituting the detailed
asymptotics of $G$ and $G_0$ to (\ref{GreenFEq}) and comparing the
terms in the left and right hand sides, we can arrive at the
equations
\begin{equation}\label{IntEqnPhi_i}
\Phi_i(x)= \Phi^0_i(x) -
\int_{{{\mathbb R}^3}}\frac{q(s)}{4\pi|x-s|}\,\Phi_i(s)\,ds\,,\quad x
\in\,{{\mathbb R}^3}\,, \quad i=1, \dots , 9\,,
\end{equation}
the first one being identical to (\ref{IntEqnPhi}).
\smallskip

\noindent {\it Step 3: completing the proof}\,\,\, The estimates
$|\Phi_i(x)| \leq c\left(1+|x|\right)^2$ easily follow from the
integral equations (\ref{IntEqnPhi_i}) and the evident estimates
of the same form for $\Phi_i^0$. Thus, we have $\Phi_i \in
{{\cal P}}^2_q$.

As is well known, the polynomials $\Phi^0_1,\, \dots ,\,\Phi^0_9$
form a basis in (unperturbed) ${{\cal P}}^2_0$. Using the equations
(\ref{IntEqnPhi_i}), it is not difficult to conclude that the same
is valid for $\Phi_1,\, \dots ,\,\Phi_9$ in the perturbed
${{\cal P}}^2_q$. \,\,\,$\square$

\subsection{Proving Theorem \ref{l3}}
Recall that we deal with the case $m=2$;
\smallskip

{\bf Necessity}\,\,\,Assume that $a \in {{\Upsilon_q}}^2$ holds. It means
that there is a function $h \in {{\cal H}}$ such that ${\left(-\Delta+q\right)}^2h=0$ in
${{\mathbb R}^3}\backslash\{a\}$ and, in the same time, ${\left(-\Delta+q\right)} h$ does not
vanish identically.

Considering $\left(-\Delta+q \right)^2h$ as a distribution on
$C^\infty_0\left({{\mathbb R}^3}\right)$, we see that it is supported at
$x=a$. Such a distribution has to be a linear combination of the
Dirac measure derivatives: $$\left(-\Delta+q \right)^2h=\sum
\limits^N_{|j|=0}\alpha_j D^j_x \delta_a= \sum
\limits^N_{|j|=0}\alpha_j (-1)^{|j|}D^j_a \delta_a\,
=:\,P_N(\nabla_a)\,\delta_a\,,$$where $P_N(\nabla_a)$ is a
differential polynomial of the order $N$ with constant
coefficients, which acts on the variable $a$. Therefore, in
accordance with (\ref{(Delta-q)^2B=delta}), the function $h$ has
to be of the form
$$h(x)= P_N(\nabla_a)\,B(x, a)+\widetilde h(x)\,, \qquad x \in {{\mathbb R}^3}\backslash \{a\}$$
with a smooth $q$-biharmonic in ${{\mathbb R}^3}$ function $\widetilde h$.

Since $$B(x,a) = c|x-a| + \rm smoother~terms$$ near $x=a$ (see
(\ref{DefB(x,y)})), the only option for $h$ to belong to ${{\cal H}}$ (to
be square-summable near $x=a$) is $\alpha_j=0$ for $|j|\geq 3$;
hence, we have $N\leq 2$, i.e.,
$$h(x)= P_2(\nabla_a)\,B(x, a)+\widetilde h(x)\,, \qquad x \in {{\mathbb R}^3}\backslash \{a\}.$$

Let us show that $\widetilde h=0$. At first, note that for $|x|\to
\infty$ the function $\widetilde h$ growths {\it not slower} than
$|x|^2$. Indeed, otherwise it grows not faster than $|x|$ and, by
the latter, the function $\varphi:={\left(-\Delta+q\right)}\widetilde h$ is
$q$-harmonic in ${{\mathbb R}^3}$ and tends to zero that is impossible by
Convention \ref{c1} \footnote{The possibility $\varphi\equiv 0$ is
also excluded by the same arguments as were used in the proof of
Lemma \ref{lemma-uniqueness}.}. On the other hand, the summand
$P_2(\nabla_a)\,B(x, a)$ may grow {\it not faster} than $|x|$,
whereas the sum $h$ belongs to ${{\cal H}}$. Hence, $\widetilde h$ must
vanish identically in ${{\mathbb R}^3}$. Thus, we get
\begin{equation}\label{representation-h}h(x)= P_2(\nabla_a)\,B(x,
a)\,, \qquad x \in {{\mathbb R}^3}\backslash \{a\}.\end{equation}

Recall that $h \in {{\cal H}}$ and tend $|x|\to \infty$. Turning to the
structure of the asymptotic of $B$ (see (\ref{asympB}),
(\ref{ansatzA}) with $y=a$) and taking into account the linear
independence of the spherical harmonics (the square brackets in
(\ref{ansatzA})), we easily conclude that the r.h.s. of
(\ref{representation-h}) can be square-summable if and only if the
coefficients at the terms, which behave as $|x|$ and
$1\over{|x|}$, vanish, i.e.,
\begin{equation}\label{PPhi=0_PPsi=0}P_2(\nabla_a)\,\Phi_i (a)\,=\,0, \quad i=1, \dots, 9; \qquad
P_2(\nabla_a)\,\Psi(a)\,=\,0\,.\end{equation}

The obtained relations for {\it $q$-harmonic} $\Phi_i$ are
nontrivial only if $P_2(\nabla_a)\not=-\Delta+q(a)$. Show that it
is the case. Indeed, otherwise, we have
$$\Phi(a)=\langle{\rm see\,} (\ref{Phi=HPsi})\rangle=-\frac{1}{6}\,P_2(\nabla)\Psi(a)=
\langle{\rm see\,} (\ref{PPhi=0_PPsi=0})\rangle\,=0\,.$$ By
(\ref{asymptoticG}), we conclude that $G(\,\cdot\,, a) \in {{\cal H}}$,
i.e., $a$ is a $s$-point of the {\it first} order, i.e., $a \in
{{\Upsilon_q}}^1$ that contradicts to the assumption.

Since $\Phi_1, \, \dots\,, \Phi_9$ constitute a basis in ${{\cal P}}^2_q$,
any $q$-harmonic polynomial $p \in {{\cal P}}^2_q$ satisfies
$P_2(\nabla)p(a)=0$. Hence, $d^2_a\left[{{\cal P}}^2_q\right] \leq 8$,
i.e., the jet degeneration at the point $a$ does occur.
\medskip

{\bf Sufficiency}\,\,\,Assume that at a point $a \in {{\mathbb
R}^3}$ the jet degeneration occurs, i.e., $d^2_a\left[{{\cal
P}}^2_q\right] <9$. Let $\cal L$ be the 10-dimensional subspace
spanned on $\Phi_1, \, \dots\,, \Phi_9, \Psi$; the degeneration
evidently implies $d^2_a[{\cal L}]<10$. By the latter, there
exists a second order differential polynomial $P_2(\nabla)$ such
that (\ref{PPhi=0_PPsi=0}) is valid. In the mean time, the same
arguments as in the item below (\ref{PPhi=0_PPsi=0}) imply
$P_2(\nabla)\not = -\Delta+q(a)$.

Defining a function $h$ by the r.h.s. of (\ref{representation-h})
and taking into account the form of the asymptotic (\ref{asympB})
and the relations (\ref{PPhi=0_PPsi=0}), it is easy to see that $h
\in {{\cal H}}$ and ${\left(-\Delta+q\right)}^2 h=0$ in ${{\mathbb R}^3}\backslash \{a\}$. Hence, $a \in
{{\Upsilon_q}}^2$ holds. \,\,\,$\square$
\medskip

In the case of $a \in {{\Upsilon_q}}^j$ for $j>2$, the proof is just more
complicated in notation. However, it exploits the same idea: a
linear combination of the relevant $q$-$j$-harmonic Green function
and its derivatives can belong to the space ${{\cal H}}$ iff the proper
analog of the relations (\ref{PPhi=0_PPsi=0}) does hold that is
equivalent to a certain jet degeneration at $x=a$.

\subsection{Radially symmetric potential}
If the potential is of the form $q=q(|x|)$, much more can be said
about structure of the $s$-point set ${{\Upsilon_q}}$. As is evident, ${{\Upsilon_q}}$
consists of spheres in ${{\mathbb R}^3}$ centered at $x=0$. Here we briefly
announce some results on this case.
\smallskip

\begin{enumerate} \item
For a fixed $l\geq 0$, the set ${{\Upsilon_q}}^l$ can be characterized as
follows. Let $\varphi_l$ be a {\it regular} solution of the radial
Schr$\ddot{\rm o}$dinger equation
\begin{align}\label{l-Schrodinger}
  -         \varphi_l ''+\frac{l(l+1)}{r^2} \,\,
  \varphi_l+q(r) \varphi_l\,=\,0\,, \qquad r>0
\end{align}
that behaves as $\varphi_l(r) \sim r^{l+1}$ near $r=0$. Note that,
in this case, the functions
$$   \frac{\varphi_l (|x|) }{|x|}  \,\, Y_l^m  \left (\frac{x }{|x|} \right )\qquad(|m|\leq l+1)  $$
are  $q$-harmonic and belong to ${{\cal P}}^l_q$. Introduce the
Kram determinants\footnote{It is worthy of noting that this
determinants appear in Darboux transform theory}
\begin{align*}
\Delta^m _l\,(r):=\,
\begin{vmatrix}
\varphi_m(r)  & \varphi_m '(r) & \cdots & \varphi_m^{(l-m)}(r)\\
\varphi_{m+1} (r) & \varphi_{m+1} '(r) & \cdots &
\varphi_{m+1}^{(l-m)}(r)\\
\cdots &  \cdots & \cdots &  \cdots\\
\varphi_{l} (r) & \varphi_{l} '(r) & \cdots &
\varphi_{l}^{(l-m)}(r)
\end{vmatrix}\,, \qquad m=0, 1, \dots ,l.
\end{align*}
Let $N_j$ be the number of the eigenvalues of the partial
Schr$\ddot{\rm o}$dinger operator $-d^2_r+\frac{j(j+1)}{r^2}+q(r)$
in $L_2(0, \infty)$ and $z(\Delta^m_l)$ the number of zeros of the
Kram determinant. The relation
\begin{align}\label{z(l)}
z(\Delta^m_l)\,=\,\sum \limits_{k=0}^{l-m} (-1)^k N_{m+k}
\end{align}
is valid. \item As we saw in section 2, $\sigma_{\rm disc}(H)\not=
\emptyset$ implies ${{\Upsilon_q}} \not=\emptyset$. Our hypothesis
is that the converse is also valid, so that the equivalence
\begin{equation}\label{hypothesis}\{\sigma_{\rm disc}(H)= \emptyset\}
\Leftrightarrow \{{{\Upsilon_q}} \not=\emptyset\}\end{equation}
holds. It can be shown that {\it for radial symmetric potentials,
if $\sigma_{\rm disc}(H)= \emptyset$ then ${{\Upsilon_q}}^k
=\emptyset$ holds for all $k\geq 1$}. This fact is established in
the following way.

One can show that $a \in {\mathbb R}^3$ is a $l$-jet degeneration
point of the $q$-harmonic polynomials iff a certain
$(l+1)\times(l+1)$-determinant $\delta_l(r)$ vanishes as $r=|a|$.
In the mean time, the determinant can be represented in the form
$$\delta_l(r)=\Delta^0_l \left(\Delta^1_l(r)\right)^2 \dots
\left(\Delta^l_l(r)\right)^2$$ that enables one to find the zeros
of $\delta_l$ by the use of (\ref{z(l)}) and, hence, to verify the
presence/absence of the jet degeneration points. \item The picture
of $s$-spheres is rather curious. For instance, in the case of
$H_\alpha=-\Delta +\alpha q$ with an interaction constant $\alpha
\geq 0$, the following scenario is possible as $\alpha$ growths:

\noindent $(a)$\,\,\,for small enough $\alpha \in [0,\alpha_1]$,
one has $\sigma_{\rm disc}(H_{\alpha})= \emptyset$ and
${{\Upsilon_q}}^k=\emptyset$ for all $k \geq 1$

\noindent $(b)$\,\,\,for $\alpha \in [\alpha_1, \alpha_2)$,
$\sigma_{\rm disc}(H_{\alpha })$ consists of a single eigenvalue
$\lambda_1^\alpha<0$, the eigenfunction $\chi_1^\alpha$ being
radially symmetric, whereas all ${{\Upsilon_q}}^k,\,k\geq 1$ are
nonempty and consist of the spheres $S_k$ of the radius $r_k$
growing to infinity as $k \to \infty$

\noindent $(c)$\,\,\,Assume that as $\alpha$ passes through
$\alpha_2$, the second negative eigenvalue
$\lambda_2^\alpha>\lambda_1^\alpha$ appears, the eigenfunction
$\chi_2^\alpha$ depending on angle variable as $Y_1(\,\cdot\,)$.
In such a case, as $\alpha\uparrow\alpha_2$, the certain spheres
$S_k, \,k\geq 2$ are blown up: $\underset{\alpha \to \alpha_2}\lim
r_k = \infty$.

In the mean time, if $\lambda_2^\alpha$ is such that
$\chi_2^\alpha$ is also radially symmetric, the picture is quite
different: all $\Upsilon_q^k,\,\,k\geq 1$ are nonempty and evolve
regularly as $\alpha$ growths.
\end{enumerate}

\subsection{Comments}
\begin{itemize} \item
Some results of the book \cite{AM} on a system of the
Schr$\ddot{\rm o}$dinger equations (\ref{l-Schrodinger}) with
different $l \geq 0$ can be interpreted in '$s$-point terms'. In
particular, the existence of the zero energy solution of the
system, which is bounded at $r=0$, leads to a certain
'non-standard' factorization of the $S$-matrix that is quite
analogous to the effects discussed in sec.1. \item Let $x=a$ be a
$s$-point of the potential $q$, and $w^f_a$ a reversing wave,
which satisfies
\begin{equation}\label{cones_a}w^f_a|_{|x-a|<|t|}\,=\,0\,.\end{equation}
The Fourier transform $\widetilde w_a (\,\cdot\,, k)=(2\pi)^{-{1
\over 2}}\int_{\mathbb R}e^{ikt}\,w_a (\,\cdot\,, t)\,dt$ obeys
$${\left(-\Delta+q\right)} \widetilde w_a=k^2\widetilde w_a\,,$$ so that $\widetilde w_a
(\,\cdot\,, k)$ is a continuous spectrum eigenfunction of the
Schr$\ddot{\rm o}$dinger operator $H$. A remarkable fact is that,
by (\ref{cones_a}), such an eigenfunction is an {\it entire}
function of the spectral parameter $k$. In due time, the existence
of such eigenfunctions was a question, which has been
affirmatively answered by R.Newton. Now we see that there exists a
rich family of entire continuous spectrum eigenfunctions
$\{\widetilde w_a (\,\cdot\,, k)\}|_{a \in {{\Upsilon_q}}}$
associated with {\it reversing finite energy waves} and
parametrized by points of the surfaces, which ${{\Upsilon_q}}$
consists of. The concrete examples demonstrate that these surfaces
may be of rather complicated shape (ovaloids, toruses, etc). The
meaning and role of such eigenfunctions in the scattering theory
are not quite clear yet. \item The important question whether the
set of the finite order $s$-points
$\cup_{m=1}^\infty{{\Upsilon_q}}^m$ exhausts ${{\Upsilon_q}}$ is
still open even in the case of radially symmetric potentials.
Also, the conjecture (\ref{hypothesis}) is not justified yet.
\end{itemize}

\bigskip

{\bf The authors:}

\bigskip
\noindent {MIKHAIL I.~BELISHEV;~~~SAINT-PETERSBURG DEPARTMENT OF
THE STEKLOV MATHEMATICAL INSTITUTE, RUSSIAN ACADEMY OF
SCIENCES;~~~belishev@pdmi.ras.ru}

\bigskip
\noindent {ALEKSEI F.~VAKULENKO;~~~SAINT-PETERSBURG DEPARTMENT OF
THE STEKLOV MATHEMATICAL INSTITUTE, RUSSIAN ACADEMY OF
SCIENCES;~~~vak@pdmi.ras.ru}~(corresponding author)

\bigskip
\noindent {\bf MSC:}~35Bxx,~35Lxx,~35P25,~47Axx

\bigskip
\noindent {\bf Key words:}~$3{\rm d}$ acoustical equation,~time
domain scattering problem, ~reversing waves, ~stop-points,
~R.Newton's factorization, ~discrete spectrum.

\end{document}